\documentclass[11pt]{article}
\usepackage{moriond,epsfig}

\def\Journal#1#2#3#4{{#1} {\bf #2}, #3 (#4)}

\def\NIMA{{\em Nucl. Instrum. Methods} A}

\def\PLB{{\em Phys. Lett.}  B}

\def\PRD{{\em Phys. Rev.} D}

\def\be{\begin{equation}}
\def\ee{\end{equation}}
\def\bea{\begin{eqnarray}}
\def\eea{\end{eqnarray}}

\begin{document}
\vspace*{4cm}
\title{Measurement of the Top Quark Mass with a Matrix Element Method in the Lepton Plus Jets Channel at CDF}

\author{ Brian Mohr (for the CDF Collaboration) }

\address{UCLA Department of Physics and Astronomy\\
475 Portola Plaza, Los Angeles, CA 90095, USA}

\maketitle\abstracts{
We present a measurement of the mass of the top quark from $p\bar{p}$ collisions
at 1.96 TeV observed with the Collider Detector at Fermilab (CDF) at the
Fermilab Tevatron Run II.  The events have the decay signature of $p\bar{p} \to t\bar{t}$ 
in the lepton plus jets channel in which at least one jet is identified as coming from a 
secondary vertex and therefore a $b$-hadron.  The largest systematic uncertainty, 
the jet energy scale (JES), is convoluted with the statistical error using an 
{\it in-situ} measurement of the hadronic $W$ boson mass.  We calculate a likelihood
for each event using leading-order $t\bar{t}$ and $W$+jets cross-sections and parameterized 
parton showering.  The final measured top quark mass and JES systematic is 
extracted from a joint likelihood of the product of individual event 
likelihoods.  From 118 events observed in 680 pb$^{-1}$ of data, we measure a
top quark mass of $174.09 \pm 2.54$ (stat+JES) $\pm 1.35$ (syst) GeV/c$^2$.
}

\section{Introduction}
The top quark mass is a fundamental parameter in the Standard Model.  Along with the mass of the $W$ boson, 
the top quark mass provides the best indication for the value of the Higgs boson mass.  This is a preliminary
measurement of the top quark mass in $p\bar{p}$ collisions at $\sqrt{s} = 1.96$ TeV with the CDF 
detector~\cite{bib:cdf} at Run II of the Fermilab Tevatron.

At the Tevatron, top quarks are mainly produced in pairs: 85\% by $q\bar{q}$ annihilation and the rest
by $gg$ fusion.  Top quarks decay before hadronization most of the time into a $b$ quark and a $W$ boson.
The $W$ bosons decay leptonically or hadronically and are used to topologically classify the final state
of the $t\bar{t}$ system.  We use events where one $W$ boson decays hadronically and
the other leptonically -- the ``lepton + jets'' channel.  Since taus have poor resolution, we include only
electrons and muons and thus sample from 30\% of the total branching ratio.

This analysis uses a matrix-element analysis technique~\cite{bib:flo} to efficiently extract 
information from the limited
number of $t\bar{t}$ events.
Each event enters the analysis with a weight derived from
the differential cross-section for $t\bar{t}$ decay.  
We take into account all the possible jet-parton assignments and integrate over unknown quantities such as
the momentum of the neutrino.  The mass of the hadronically decaying $W$ boson
is measured and constrained with the current world average of 80.4 GeV/c$^2$ to measure the jet energy
scale (JES) of the events.  The jet energy scale is the largest source of systematic uncertainty in this
analysis.  The final top quark mass is extracted from a simultaneous fit of the top quark mass and JES.

\section{Data Sample \& Event Selection}
\label{sec:acceptance}
Events from the lepton + jets decay channel are selected requiring a single, high-transverse energy, 
well-isolated lepton; large missing transverse energy; and exactly four, central, high-transverse energy jets.  
(Two jets originate from the $b$ quarks and two from the hadronically decaying $W$
boson.)  Of these jets, we require at least one to be ``$b$-tagged,'' identified as originating 
from a secondary vertex
and thus the decay of a long lived $b$ hadron.  The primary
vertex is the one from which tracks associated with the lepton emerge.  The secondary vertex tag identifies
tracks originating from a vertex displaced from the primary vertex and associated with a jet.  To reduce the
amount of non-$W$ background, we further require the leading jet and missing transverse energy not be collinear
in the transverse plan for the lowest values of missing transverse energy passing our selection.  
Table~\ref{tab:acc} outlines this event selection.

\begin{table}[h]
\caption{Event Selection.\label{tab:acc}}
\vspace{0.4cm}
\begin{center}
\begin{tabular}{|l|c|}
\hline
lepton         &  $E_T > 20$ GeV ($e$), $p_T  > 20$ GeV/c ($\mu$) \\
jets           &  $E_T > 15$ GeV, $|\eta| < 2.0$ \\
missing $E_T$  &  missing $E_T > 20$ GeV \\
$b$-tag        &  $\ge 1$ jet coming from secondary vertex \\
QCD veto       &  $0.5 < \Delta\phi < 2.5$ (missing $E_T < 30$ GeV) \\
\hline
\end{tabular}
\end{center}
\end{table}

\section{Method}
We write a likelihood for each event by combining a signal probability with a background probability.
This method was first used to measure the top mass by the D\O$\;$ collaboration during
Run I~\cite{bib:flo}.  The likelihood is minimized for three variables: the top quark mass, 
the jet energy scale
(JES), and the fraction of events consistent with our signal hypothesis ($C_s$), where 
$\vec{x}$ are the measured quantities:
\begin{equation}
  \mathcal{L}(M_{top},JES,C_s;\vec{x}) \propto \prod_{i=1}^N [C_s P_{t\bar{t}}(\vec{x};M_{top},JES) + (1 - C_s) 
    P_{W+\mathrm{jets}}(\vec{x};JES)] \mathrm{.}
\end{equation}
We first minimize the likelihood for $C_s$ with {\tt MINUIT} and
then perform a two dimensional fit to extract $M_{top}$ and $JES$.  The signal probability, $P_{t\bar{t}}$,
indicates how well an event describes leading order $t\bar{t}$ pair production and decay, and the
background probability, $P_{W+\mathrm{jets}}$, indicates how well an event describes the 
largest contributing background
process, a leptonically decaying $W$ boson plus extra jets.  
These probabilities are calculated by integrating over the parton differential cross-section,
where the measured quantities are input to detector resolution functions, $W(\vec{x},\vec{y})$,
used to transfer the parton quantities, $\vec{y}$,
to measured quantities, and we also
include parton distribution functions, $f(\tilde{q_i})$, for $p\bar{p}$ collisions:
\begin{equation} 
  P(\vec{x})=\frac{1}{\sigma_{t\bar t}}\int d\sigma(\vec{y}) W(\vec{x},\vec{y}) f(\tilde{q_1}) 
  f(\tilde{q_2}) d\tilde{q_1} d\tilde{q_2} \mathrm{.}
  \label{eqn:prob}
\end{equation}
We use an analytical form of the leading
order matrix-element for $q\bar{q} \to t\bar{t}$ in the signal probability and the sum of the 
$W$ + 4 jets matrix-elements of the {\tt VECBOS} Monte Carlo generator in the background probability.

Our input measured quantities are the momentum of the lepton and the angles and energies of the jets.
Regarding detector resolution, we consider the momentum of the electron or muon and the jet angles
to be well measured.  Thus, the detector resolution in Equation~\ref{eqn:prob} corresponds
to the jet energy resolution, which is modeled from Monte Carlo~\cite{bib:pyt}~\cite{bib:hwg} 
using a ``transfer function,''
a mapping between jet energies and parton energies, $W_{jet}(\vec{x},\vec{y})$.  
Signal and background probabilities are summed over all possible permutations of jet and parton 
combinations, and the 
signal probability also considers different possible values of the transverse momentum of the
$t\bar{t}$ system.

We are sensitive to the jet energy scale through the mass of hadronically decaying $W$ boson by constraining
it with the world average $W$ boson mass of 80.4 GeV/c$^2$.  We define the JES as a multiplicative scale 
factor applied to the energies of the
two jets selected as the daughters of the $W$ boson decay:
\begin{equation}
E_{jet} = E_{jet}^{MC} / JES \mathrm{.}
\label{eqn:JES}
\end{equation}
The constraint comes from integrating over the event mass set by the daughters of the $W$ boson and the 
JES in a Breit-Wigner using the world average value as the pole mass.
We assume the JES determined for $W$-jets also applies to $b$ jets and assign a systematic uncertainty
for the difference between the $W$ and $b$ jet energy scale.

\section{Systematic Uncertainties}
\begin{table}[h]
\caption{Sources and Values of Systematic Error.\label{tab:syst}}
\vspace{0.4cm}
\begin{center}
\begin{tabular}{|l|c|}
\hline
Source of systematic uncertainty         &  Magnitude (GeV/c$^2$)    \\
\hline
Residual jet energy scale                &  0.42  \\
b-jet energy scale                       &  0.60  \\
Generator                                &  0.19  \\
Initial state radiation                  &  0.72  \\
Final state radiation                    &  0.76  \\
b-tag $E_T$ dependence                   &  0.31  \\
Background composition                   &  0.21  \\
Parton distribution functions            &  0.12  \\
Monte Carlo statistics                   &  0.04  \\
\hline
Total                                    &  1.35  \\
\hline
\end{tabular}
\end{center}
\end{table}

Table~\ref{tab:syst} lists the systematic uncertainties estimated from various Monte Carlo samples
and re-weighting techniques.  To first order, we fit out the JES systematic of our likelihood, but we
also apply a residual systematic to cover an higher order effects, such as variations in the expected
$\eta$ or $p_T$ distribution.  This higher order effect is estimated by shifting the input jet energies
up and down one sigma as defined by the CDF Jet Energy and Resolution group \cite{bib:jer}.  The
generator systematic uncertainty takes into account differences in the fragmentation and showering
by comparing two different Monte Carlo models ({\tt PYTHIA} and
{\tt HERWIG}).  Possible biases originating from differences in the amount of initial- and final-state
radiation between data and Monte Carlo are estimated using {\tt PYTHIA} samples generator with lesser
or greater amounts of radiation.  The uncertainty
on the parton distribution functions is evaluated as the sum in quadrature of the difference between
{\tt MRST} and {\tt CTEQ} parton distribution functions; {\tt MRST} with $\Lambda_{QCD} = 228$~MeV
and $\Lambda_{QCD} = 300$ MeV; and the variation of the 20 {\tt CTEQ6} eigenvectors.  Systematic
effects from the dependence of the $b$-tagging on $p_T$ are evaluated by changing the dependence by
one sigma.  The background composition and modeling systematic error is the sum in quadrature of 
the largest variation when fluctuating the contribution of each individual background sample
by 100\%; the change in the signal fraction by $\pm 10$\%; and the largest variation due to changing the
$Q^2$ scale in $W$ + jet production.  To understand the effects of limited statistics in the background
sample, we divide the smallest sample (non-$W$ background) in half and compare the results using each
half separately.  We repeat this procedure several times and histogram the difference.  We take
one-half the RMS of the distribution as a systematic error.

\section{Results}
\begin{figure}
\begin{center}
\psfig{figure=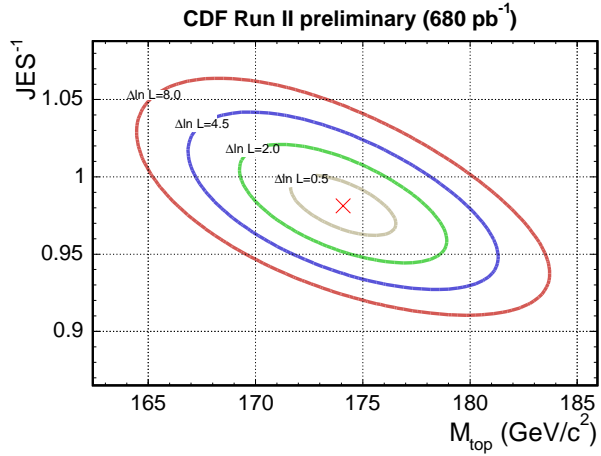,height=2.4in}
\caption{Likelihood after minimization of $C_s$ parameter.
\label{fig:lik}}
\end{center}
\end{figure}

The output likelihood is a simultaneous fit to three parameters: top quark mass, jet energy scale (JES) 
and signal
fraction ($C_s$).  We do not use any prior knowledge of JES or $C_s$ in the extraction of the top mass.  The
data used in the analysis corresponds to that collected in the period between March 2002 and September
2005, a total integrated luminosity of 680 pb$^{-1}$, where 118 events pass event selection.  Figure~\ref{fig:lik}
shows the fit to this data after minimization for $C_s$ as a function of $M_{top}$ and $JES$ with different
$\Delta \mathrm{ln} \mathcal{L}$ contours.  Our measurement is:
\begin{equation}
M_{top} =  174.09 \pm 2.54 \; \mathrm{ (stat+JES) } \pm 1.35 \; \mathrm{ (syst)} \; \mathrm{ GeV/c}^2
\end{equation}

\end{document}